\documentclass[reprint,amsmath,amssymb]{revtex4-1}
\usepackage{graphicx}%
\usepackage{amsmath}
\usepackage{dcolumn}%
\usepackage{bm}%
\usepackage{xcolor}
\usepackage[utf8x]{inputenc}
\usepackage{hyperref}

\begin{document}

\title{Positron Acceleration in an Elongated Bubble Regime}

\author{Tianhong Wang$^1$, Vladimir  Khudik$^2$, and Gennady Shvets$^1$}
 \affiliation{$^1$School of Applied and Engineering Physics, Cornell University, Ithaca, New York 14850, USA.\\$^2$Departmentx of Physics and Institute for Fusion Studies, The University of Texas at Austin, Austin, Texas 78712, USA.}

\date{\today}%

\begin{abstract}
A new concept is proposed for accelerating positrons in a nonlinear plasma wakefield accelerator. By loading the wakefield (back of the plasma bubble) with a short electron bunch, an extended area of excessive plasma electron accumulation is created after the first bubble, resulting in a favorable region with simultaneous focusing and accelerating fields for positrons. Scaling laws for optimized loading parameters are obtained through extensive parameters scans. Owing to the good quality of the focusing field, positron acceleration with emittance preservation can be achieved in this new regime and it has been demonstrated in the three-dimensional particle-in-cell simulations.

\end{abstract}

\maketitle

{\it Introduction and Motivation.} 

Plasma-based accelerators are able to sustain a large acceleration gradient: $>{\rm GeV/m}$ without being subject to the material electrical breakdown, providing a new pathway to more compact and affordable colliders~\cite{Colliders1, Colliders2,RoadMap} compared with the next-generation conventional lepton colliders~\cite{CLIC,InternationalLC}.  In the plasma accelerators, a trailing bunch of electrons is accelerated in the plasma wakefield created either by a high-intensity laser pulse for a laser wakefield accelerator (LWFA)~\cite{LWFA_1,LWFA_2,LWFA_3}, or by an ultrarelativistic charged particle beam for a plasma wakefield accelerator (PWFA)~\cite{PWFA_1}.  Tremendous progress has been made in the past few decades: multi-GeV electron accelerations at a single stage have been demonstrated over a plasma length of a few centimeters to several tens of centimeters, both in the LWFA regime~\cite{GeV_0,GeV_1,GeV_2,GeV_3,GeV_2b} and PWFA regime~\cite{PWFA_GeV_1,PWFA_GeV_2,PWFA_GeV_3}. 

For the sake of acceleration efficiency and emittance preservation, electrons are often accelerated in the nonlinear blowout, or the so-called "bubble" regime where high-gradient accelerating field and linear focusing field are both provided~\cite{PWFA_Field_1991,LWFA_Field_2002,Mapping_Field_2016}. However, their counterpart: positrons are difficult to accelerate in the same regime due to the lack of a suitable focusing field created by the excess of negative charges: electrons. In the bubble regime, massive electron accumulation only occurs at the end of the bubble, forming a density spike that is much shorter than $k_p^{-1}$, unsuitable for positron acceleration~\cite{Lotov_Positron}.

The key to accelerating positrons in a nonlinear regime is to produce a considerable large overlap between the accelerating phase and the focusing phase. Several groups have proposed approaches to extend the volume of electron accumulation, by using a positron driver beam~\cite{Nature_Positron_2015}, a hollow electron driver beam~\cite{Hollow_Positron_2015}, or finite-size plasma column~\cite{Column_Positron_2019,Column_Positron_2020}. Each of them has its drawbacks. i)  In the self-loaded wake created by a positron driver beam, the focusing field is nonlinear, therefore, detrimental for emittance preservation. ii) Hollow electron beam imposes challenges on the generation and preservation of such beam. Moreover, since the driver beam and witness positron bunch are kept at the same longitudinal location~\cite{Hollow_Positron_2015}, the transformer ratio (R.T.) of this concept is limited to be $<1$.  iii) Finite-size plasma column limits the parameter space for both the driver beam and the plasma due to the existence of self-field of the driver beam. Furthermore, the transverse focusing field is not linear therefore, a Gaussian witness bunch is not matched~\cite{Column_Positron_2019}. In addition, creating and maintaining a long plasma column with a sharp boundary would be experimentally challenging.

\begin{figure}[t!]
\centering
  \includegraphics[width=0.95\columnwidth]{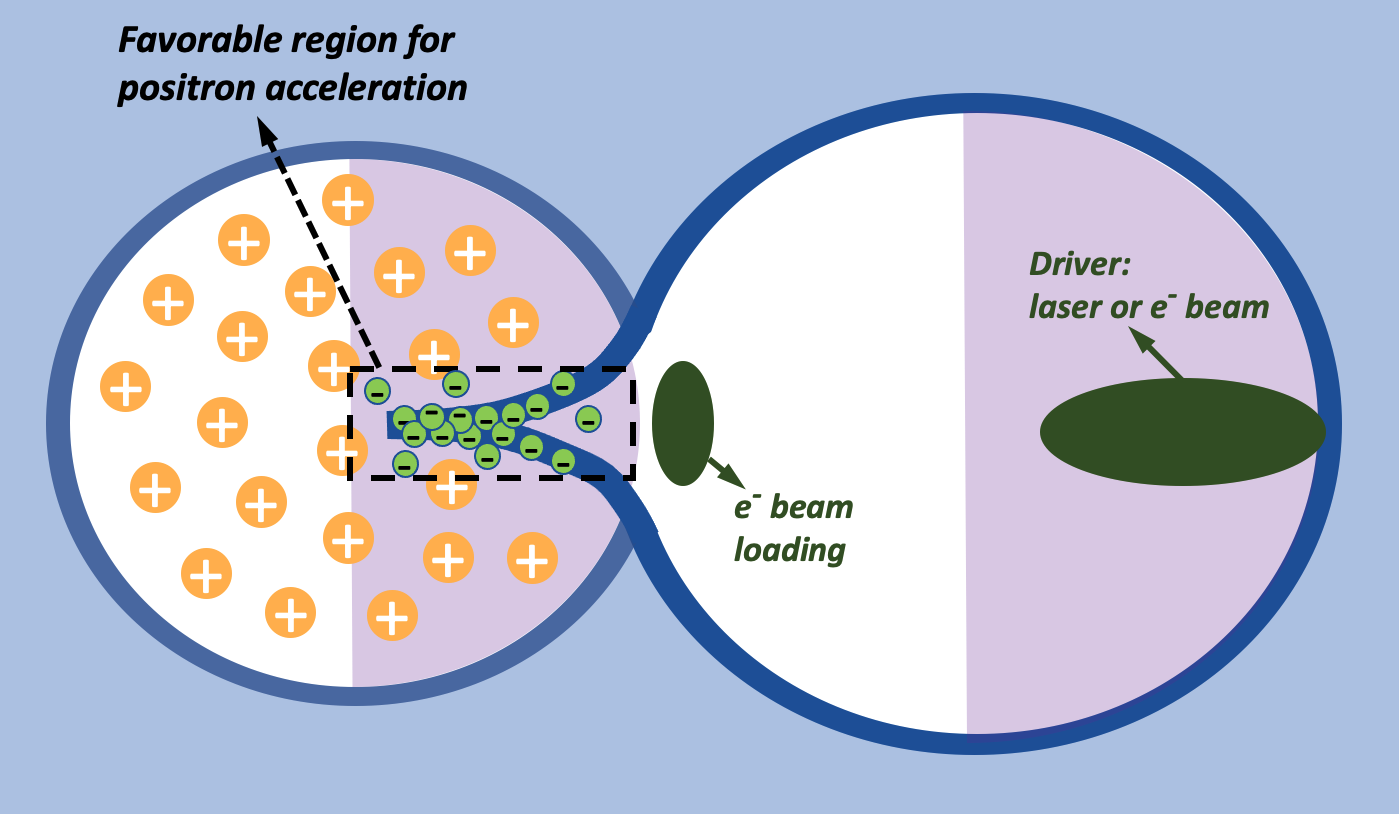} \\
  \caption{Schematic of the ELBA concept. Orange particles represent the ions. Green particles represent electrons. Dark-blue curves represent the plasma electron flows along the bubble boundary. Pink shadings show the accelerating phases for positrons.}\label{Fig1}
\end{figure}

In this letter, we present a new concept of creating a region favorable for positron acceleration in a fully blowout bubble by loading the back of the bubble with a short electron bunch. As shown in Fig.~\ref{Fig1}, the electron loading rapidly decelerates the relativistic plasma flows at the back of the bubble and extends them into the first half (accelerating phase) of the second bubble. This extended electron accumulation with high charge density produces a strong focusing field for positrons. Since the back of the bubble is elongated, we call this regime "elongated bubble acceleration" (ELBA).  As a starting point, we show a three-dimensional (3D) simulation of the ELBA regime and present the characteristics of the fields which are ideal for the acceleration and quality preservation of the positrons.

{\it Characteristics of ELBA Regime.} 
\begin{figure}[t!]
\centering
  \includegraphics[width=0.95\columnwidth]{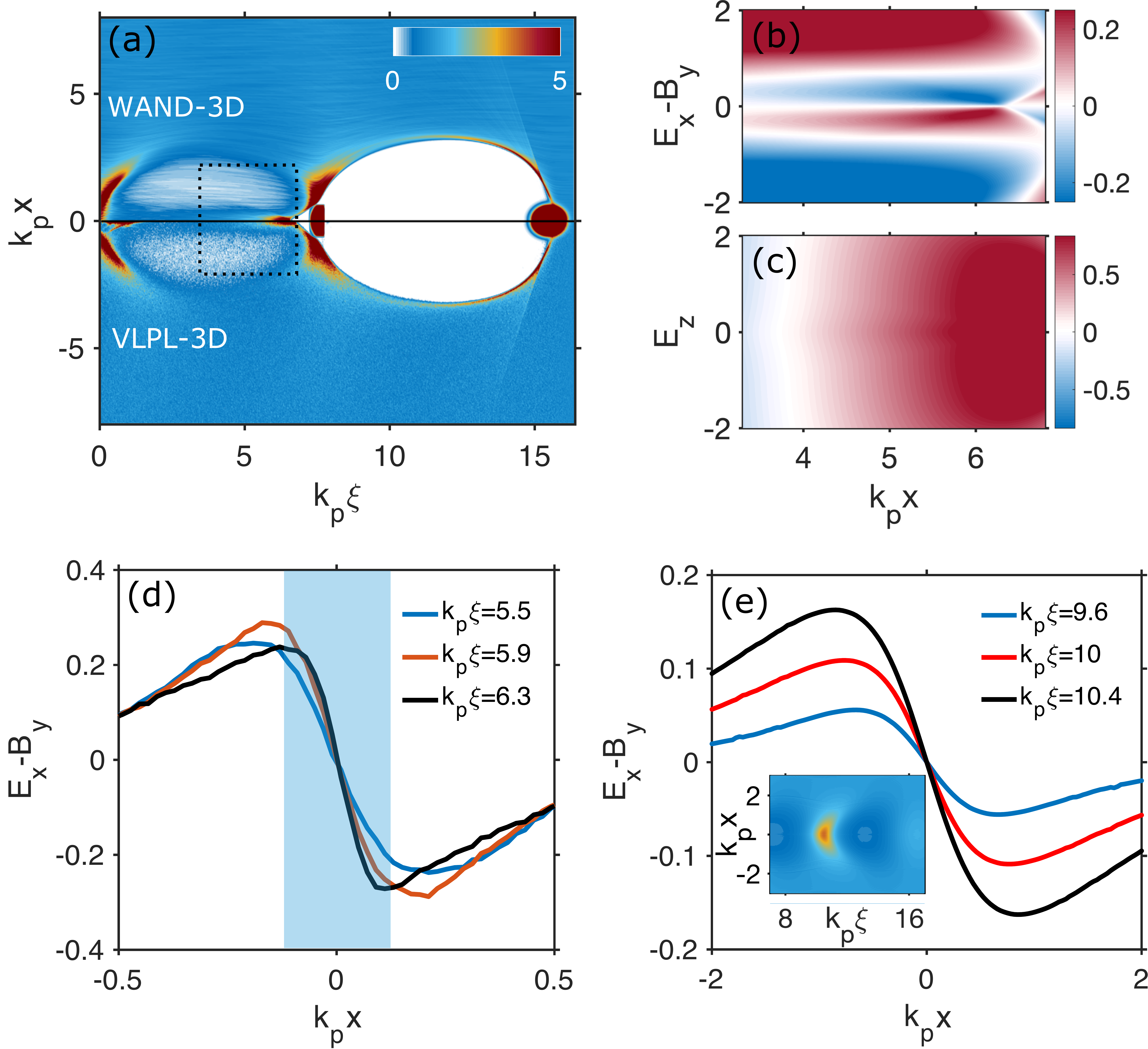} \\
  \caption{Example setup of ELBA scheme in 3D simulations.  (a) Side-by-side comparison of simulations results from WAND-PIC (upper half-plane: x>0) and VLPL-3D (lower half-plane: x<0). (b, c) Transverse focusing field $F_x=E_x-B_y$ and accelerating field $E_z$, plotted from the dashed region in (a). (d) Lineout of focusing field $F_x$ at three different locations: $k_p\xi=5.5$, $k_p\xi=5.9$, and $k_p\xi=6.3$, where $\xi=z-ct$ is the longitudinal co-moving coordinate. The blue shaded rectangle indicates the liner focusing region. (e) Lineout of focusing field $F_x$ at three different locations: $k_p\xi=9.6$, $k_p\xi=10$, and $k_p\xi=10.4$ in a quasi-linear regime. The inset: plasma density in the quasi-linear regime. Simulation parameters for the QL regime: plasma density $n_0=2.2\times10^{16}cm^{-3}$. Driver charge: $q=1.5nC$, transverse size (FWHM) $\delta=89.5\mu m$, duration (FWHM) $\tau=300fs$.}\label{Fig2}
\end{figure}

ELBA concept is examined in a fully blowout regime driven by a short electron bunch with a normalized charge $Q>1$~\cite{stupakov_2016,My_Driver_2017}, where $Q=\omega_p^3q/(4\pi en_0c^3)$, $q$ is the bunch charge, $\omega_p = \sqrt{4\pi e^2n_0/m}$ is the plasma frequency with plasma density $n_0$, $c$ is the speed of light, and $-e$ and $m$ are the electron charge and mass, respectively. A driver electron bunch with $q_1=1nC$ ($Q_1=2.4$) and duration $\tau_1=13fs$ propagates in a plasma with density $n_0=5\times10^{17}cm^{-3}$. A second electron bunch with $q_2=0.5nC$ ($Q_2=1.2$) and duration $\tau_2=6.5fs$ trails behind by a distance $L=56.4\mu m$. Fig.~\ref{Fig2}~(a) shows the side-by-side comparison of the elongated bubbles obtained from a quasi-static simulation done by WAND-PIC~\cite{WAND2020} and a full particle-in-cell simulation done by VLPL-3D~\cite{VLPL1999,VLPL2020}, excellent agreement is observed. In both simulations, we included a initial plasma temperature $T_e=72eV$ and the effect of temperature will be discussed later.  Due to the beam-loading effect, the second electron bunch $q_2$ elongates the back of the first bubble and extends the electron density filament all the way across the shallower, second bubble. This electron filament covers the whole positron accelerating phase in the second bubble and provides focusing for the positrons.  Fig.~\ref{Fig2}~(b), (c) show the accelerating field $eE_z/m\omega_pc$ and the focusing field $F_x=e(E_x-B_y)/m\omega_pc$, respectively, from the dashed region in Fig.~\ref{Fig1}~(a). The length and width of the positron favorable region, at where accelerating field and focusing field overlap, are $2k_p^{-1}$ and $1.2k_p^{-1}$.  For the consideration of emittance preservation~\cite{FACETII_2018}, a focusing field which is linear in $r$: $\partial F_r/\partial r=const$ and independent of $\xi$: $\partial F_r/\partial \xi=0$ is highly valued, where $\xi=z-ct$, since it minimizes the betatron decoherency~\cite{Mapping_Field_2016,Emittance_Xu_2012,Emittance_Mehrling_2012} of different bunch slices. Here we plot the $F_x(x)$ at three different longitudinal locations in Fig.~\ref{Fig2}~(d): $k_p\xi=5.5$, $5.9$, and $6.3$. We found that the focusing field is homogeneous within $k_p^{-1}$ and there is a linear focusing region at the center of focusing fields. The transverse size of linear focusing region is $0.24k_p^{-1}$. 

A quasi-linear (QL) regime has been also proposed since it has a higher gradient than a linear regime while still providing a viable focusing structure~\cite{RoadMap,QuasiLinear_2017,QL_An_2019}. The advantage of the ELBA regime is evident compared with the QL regime. A simulation with the same total electron charge $1.5nC$ but in a less dense plasma is conducted and the results are plotted in Fig.~\ref{Fig2}~(e). The length of the positron favorable region (at where accelerating and focusing phases overlap) in the QL regime is $0.8k_p^{-1}$. Although one can always find a linear focusing field at the vicinity of the axis ($r=0$), the focusing gradient is not independent of $\xi$, as evidenced by  Fig.~\ref{Fig2}~(e). In terms of accelerating gradient $F_z$, the ELBA regime possesses $7.6$ times higher accelerating gradient for positrons compared with the QL regime.

{\it Effect of Plasma Temperature} 

In a cold plasma, it is well-known that the peak electron density becomes singular when the electric field amplitude approaches the wave-breaking field $E_{WB}=\sqrt{2(\gamma_\varphi-1)}E_0$~\cite{WB_JETP_1956,WB_Dawson_1959}, where $\gamma_\varphi=1/\sqrt{1-\beta_\varphi^2}$, $v_\varphi=c\beta_\varphi$ is the plasma wave phase velocity, and $E_0=m_e\omega_pc/e$ is the nonrelativistic wave-breaking field~\cite{WB_Dawson_1959}. From the warm fluid model~\cite{Esarey_Temp_2004,Esarey_Temp_2005}, an initially cold collisionless plasma remains cold. Therefore, in a nonlinear PWFA simulation that starts with a cold plasma, the exact profile and height of the density spike at the boundary of the bubble cannot be accurately simulated due to its singular nature. Accordingly, in the ELBA setup, we found that the peak density of the electron filament formed behind the bubble increases with the resolution if a cold plasma is used. However, in a warm plasma, the catastrophic density compression is prevented by the thermal pressure~\cite{WB_Warm_Coffey,WB_Warm_Schroeder,WB_Warm_Schroeder2}. It is found that the maximum density compression in the 1D warm plasma is~\cite{WB_Warm_Schroeder}: $n_{max}/n_0=(3\theta)^{-1}+1/2$, where $\theta=k_BT_e/mc^2$ and $k_B$ stands for the Boltzmann constant, making the width of density spike $\propto T$. In the ELBA simulations, it is necessary to use a warm plasma not only because it produces a realistic/non-singular density profile, but also because it widens the electron filaments and produces a wide linear focusing region, as shown in Fig.~\ref{Fig2}~(d). In contrast, cold plasma cannot produce the same structure due to the existence of infinite electron density (and gradient). The steplike focusing field reported in ~\cite{Column_Positron_2019,Column_Positron_2020} could be explained by this phenomenon. We found that tens of $eV$ initial electron temperature in ELBA setup is enough to create a linear focusing region with size $0.1\sim0.2k_p^{-1}$. Such warm plasma is common in laboratories when plasma is created through laser ionization\cite{Channel_ionization,Channel_ionization2}. For more details about the effect of temperature, please see the Sec. I of the Supplemental Material.

{\it Phenomenological Model and Scaling.} 

\begin{figure}[t!]
\centering
\includegraphics[width=0.95\columnwidth]{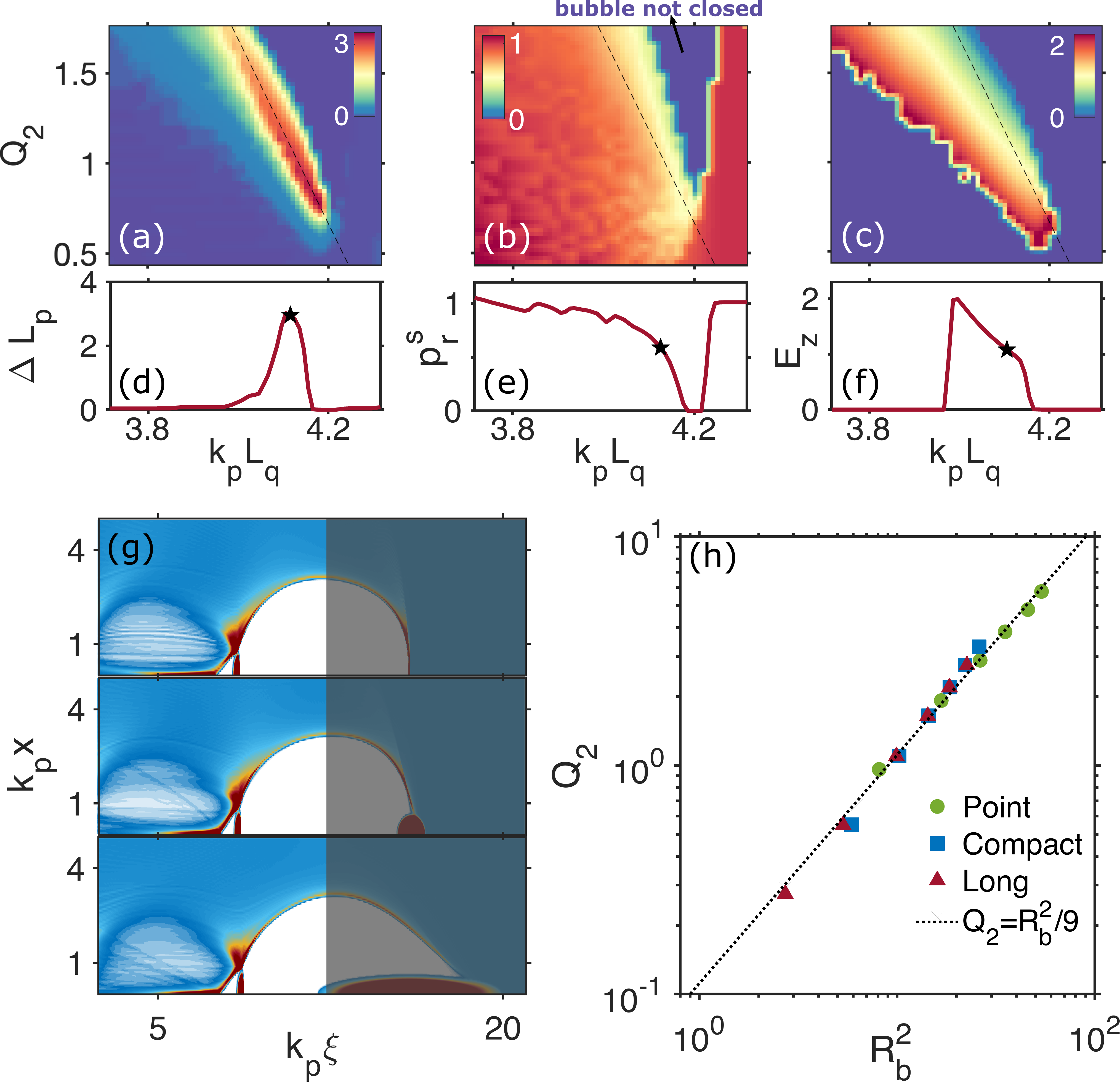} \\
\caption{Simulation scan and scaling law. (a - f) Results of simulation scan with various loading charges $Q_2$ and bunch positions $L_q$: the distance between $Q_2$ and bubble center. In the simulation scan, $Q_1=2.4$ and the shape is fixed, so is the bubble radius. (a) Length of positron favorable region: $\Delta L_{p}$ as a function of $Q_2$ and $L_q$. (b) Transverse momentum of plasma sheath: $p_r^{s}$ on axis at the end of the first bubble. (c) Peak accelerating field: $E_z$ in the positron favorable area. (d - f) Horizontal cuts of the sub-figures (a - c) at $Q_2=1.1$. The black stars in (d - f) represent the optimal locations for $Q_2$. (g) Three different bubbles with same radius: $R_b=3.2k_p^{-1}$, driven by three different driver bunches, from top to bottom: point driver ($Q_1=1.8$), compact driver ($Q_1=2.4$), long driver ($Q_1=4.0$). (e) The optimal charge $Q_2$ for different bubble radius $R_b$ and driver profiles.}\label{Fig3}
\end{figure}

The key to creating an elongated positron accelerating structure is to load the back of the bubble properly, i.e., by placing the right amount of charge at the appropriate location to decelerate the relativistic plasma flows. A small loading charge creates insufficient electron deceleration, while a big loading charge will rebound the plasma electrons and opens up the bubble. To understand what would be the best loading, a parameter scan is performed by scanning different values of $Q_2$ and different distances of $Q_2$ from the bubble center: $L_q$. Fig.~\ref{Fig3}~(a-f) show the results of the parameter scan (1024 simulations in total) when the driver charge is fixed to be $Q_1=2.4$ and the resulting bubble radius $R_b=3.2k_p^{-1}$. Fig.~\ref{Fig3}~(a) shows the overlapping length of the focusing phase and accelerating phase: $\Delta L_{p}$, and Fig.~\ref{Fig3}~(d) is the horizontal cut of Fig.~\ref{Fig3}~(a) at $Q_2=1.1$. The maximum overlapping length $\Delta L_{p}^{max}\approx \lambda_p/2$, much bigger than the length of such area in a linear regime: $\lambda_p/4$. We also notice that there is an "optimal band" in the parameter space, and the width of this band is about $0.1k_p^{-1}$, which represents the tolerance for placing $Q_2$. There also exists a lower bound for the value of $Q_2$ below which no positron accelerating structure is created due to insufficient loading.  Fig.~\ref{Fig3}~(a) also illustrates that, as long as the $Q_2$ is not destroying the general bubble structure, a bigger $Q_2$ may be placed further away from the back of the bubble compared with a smaller $Q_2$ however, they should provide the same amount of deceleration to the plasma trajectories as demonstrated by Fig.~\ref{Fig3}~(b).  Fig.~\ref{Fig3}~(b, e) shows the transverse momentum of the trajectories in the plasma sheath: $p_{r}^{s}$ at the end of the first bubble. An insufficient loading (either $Q_2$ is too small or isn't close enough to the back) will provide poor deceleration to the plasma trajectories, therefore trajectories cross the axis and create a short density spike. However, too much deceleration ($p_{r}^{s}\rightarrow0$) will delay the crossing of the trajectories, or even rebound the trajectories (the purple region in Fig.~\ref{Fig3}~(b)). Combining Fig.~\ref{Fig3}~(a) and (b), we found that the momentum in the optimal band is around $0.6$, while an unloaded bubble would have $p_{r}^{s}>1$. Nevertheless, the presence of $Q_2$ decreases the accelerating field experienced by the positrons, as shown in Fig.~\ref{Fig3}~(c) and (f). The wave-breaking field can still exceed $E_z$ ($eE_z/m\omega_pc>1$) if we choose the right $Q_2$ and $L_q$ in the optimal band. Same parameter scans are conducted for the other two bubble radii: $R_b=3.8k_p^{-1}$ and $R_b=4.3k_p^{-1}$ (please see the Sec. II of the Supplemental Material), and we found the following formula which can be used to describe the optimal band:
\begin{eqnarray}\label{eq_OptiBand}
\frac{\partial Q_2}{\partial L_q}= -1.6 R_b,\quad
\end{eqnarray}
where $L_q$ and $R_b$ are both normalized by $k_p$.

And Eq.~\ref{eq_OptiBand} is plotted in Fig.~\ref{Fig3} as dashed lines. Among the optimal band, we found that there is an optimal value of $Q_2$ such that it creates a best positron accelerating structure: a big $Q_2$ may deform the bubble back structure and decrease the overlapping length $\Delta L_{p}$ and a small $Q_2$ creates less uniform electron accumulation density. For example, in Fig.~\ref{Fig2} and~\ref{Fig3}, the optimal charge for a bubble radius $R_b=3.2$ is $Q_2=1.1$. Furthermore, the optimal loading charge $Q_2$ needed to create such structure only depends on the size of the bubble $R_b$ regardless of the driver's shape as long as the bubble is a fully blowout.  Fig.~\ref{Fig3}~(g) shows three different bubbles driven by a point charge ($Q_{1}=1.8$), compact bunch ($Q_{1}=2.4$), and a long bunch ($Q_{1}=4.0$), respectively. The optimal loading bunches used in three simulations are the same: $Q_{2}=1.1$, since these three bubbles have identical structures after the center of the bubble at where $k_pr=R_b=3.2$.  Further simulations with different $R_b$'s and different driver profiles show that the scaling law:
\begin{eqnarray}\label{eq_Scaling}
Q_2\approx R_b^2/9
\end{eqnarray}
holds in most of the cases when $R_b\gg1$ and driver's transverse size $\ll1$, as shown in Fig.~\ref{Fig3}~(h). We also notice that a similar scaling holds when an intense laser pulse plays the role of diver, however the constant $C=R_b^2/Q_2$ may vary slightly (normally within $20\%$) depending on the size and shape of the laser pulse. The scaling law Eq.~\ref{eq_Scaling} can be derived from the widely-used differential equation which describes the boundary of a blowout bubble:~\cite{Lu_bubble_2006,Yi_bubble_2013}
\begin{eqnarray}\label{eq_Lu}
r_b\frac{\partial^2r_b}{\partial\xi^2}+2\Big(\frac{\partial r_b}{\partial\xi}\Big)^2+1=4\frac{\lambda(\xi)}{r_b^2},
\end{eqnarray}
where $r_b(\xi)$ is the bubble radius as a function of $\xi$, and $\lambda(\xi)$ is the normalized charge per unit length~\cite{Lu_bubble_2006}. In our case, we assume $\lambda(\xi)=Q_2\delta(\xi-L_q)/2$, where $\delta(\xi)$ is the Dirac-delta function. It's easy to see that if we normalize both $r_b$ and $\xi$ by $R_b$, Eq.~\ref{eq_Lu} becomes a normalized equation with an universal solution if only if $Q_2/R_b^2=const$.  

{\it 3D Simulation.} 

\begin{figure}[t!]
\centering
  \includegraphics[width=0.98\columnwidth]{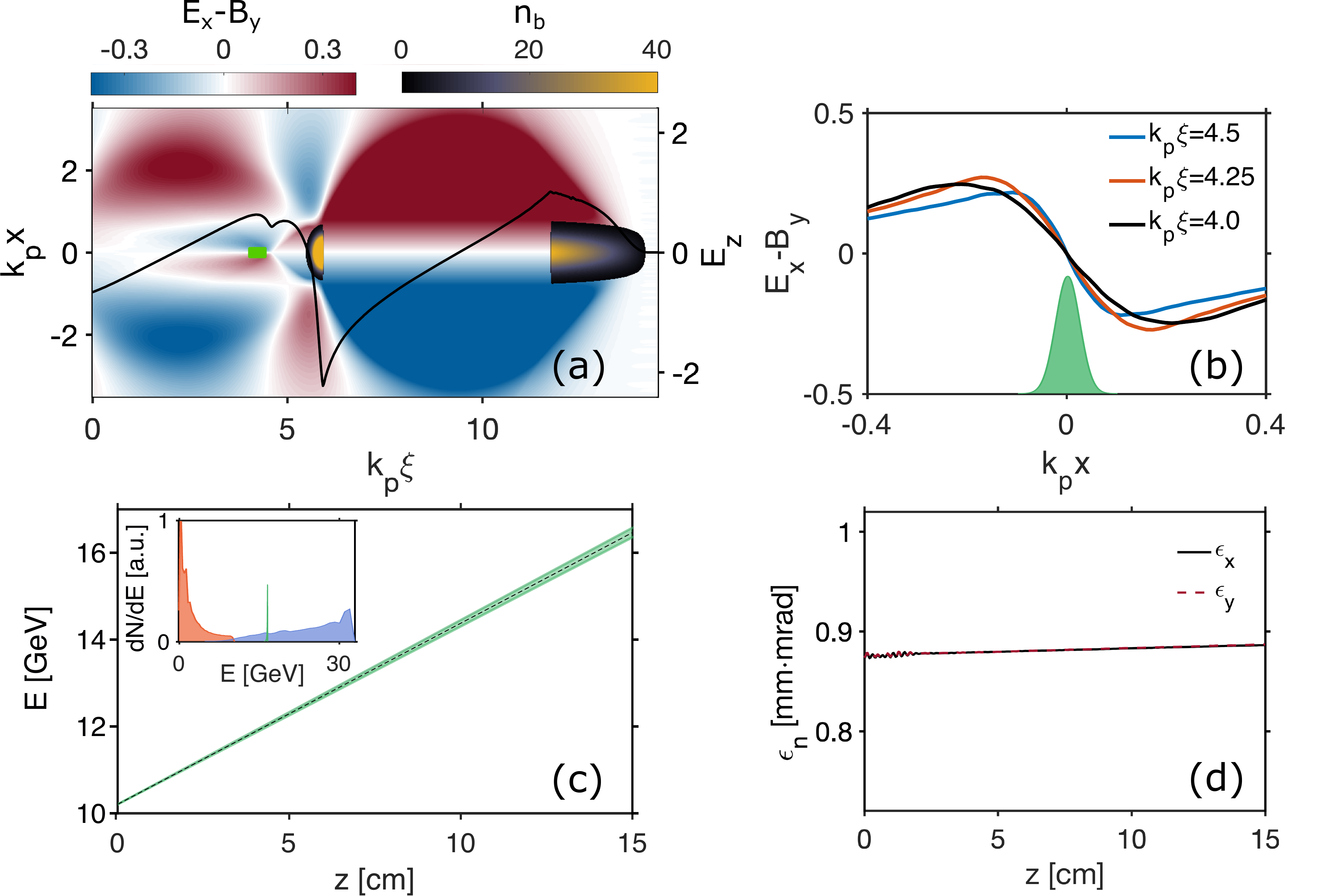} \\
  \caption{Full 3D Simulation of ELBA.  (a) Transverse field ($E_x-B_y$, red-blue colormap) and longitudinal field ($E_z$, solid line) at the initial moment $z=0 cm$. The densities of driver bunch $q_1$ and loading bunch $q_2$ are shown by the black-yellow colormap. Green dots represent positrons. (b) Focusing field $F_x=E_x-B_y$ at three different locations: $k_p\xi=4.0$, $k_p\xi=4.25$, and $k_p\xi=4.5$. The green shading shows the location and the transverse shape of positron bunch. (c) The energy evolution and energy spread of the positron bunch. The inset: final energy spectrum of the driver bunch: red, loading bunch: blue, and positron bunch: green. (d) Normalized transverse emittance of the positron bunch. Simulation parameters:  the driver bunch has $q_1=0.85nC$ ($Q_1=2.0$), transverse size (FWHM) $\delta_1=5\mu m$, duration (FWHM) $\tau_1=31fs$ and a linear longitudinal current profile. The loading electron bunch has $q_2=0.34nC$ ($Q_2=0.8$), transverse size (FWHM) $\delta_2=3.75\mu m$, duration $\tau_2=4.2fs$, and a half Gaussian longitudinal current profile. The two electron bunches are separated by a distance $L=53.1\mu m$. Initial plasma temperature $T_e=72eV$.}\label{Fig4}
\end{figure}

We conduct a full 3D simulation to examine the acceleration of a positron bunch and its emittance preservation. As shown in Fig.~\ref{Fig4}~(a), we choose a longer driver bunch compared with the driver we used in Fig.~\ref{Fig2}~(a), in order to increase the transformer ratio.  Both the driver bunch $q_1=0.85nC$, loading bunch $q_2=0.34nC$, and the positron bunch are launched with initial energy $10.2GeV$ and propagate in a plasma with $n_0=5\times10^{17}cm^{-3}$ (for more parameters, please see the caption of Fig.~\ref{Fig4}). Figure~\ref{Fig4}~(b) shows that this setup generates a positron favorable region ($\partial F_r/\partial r=const$ and $\partial F_r/\partial\xi\approx0$) with width $\approx0.3 k_p^{-1}$ and length $\approx0.5k_p^{-1}$. The positron bunch is a Gaussian beam with negligible charge, its transverse size (FWHM) $\delta_p$= $0.06k_p^{-1}$ and duration (FWHM) $\tau_p=0.16 k_p^{-1}$. 

After a propagation distance $z=15cm$, the positron bunch is accelerated to average energy $16.5GeV$ with energy spread $\Delta E/E=1.4\%$ until the accelerating/focusing structure starts to degrade due to the depletion of the driver bunch. The average energy loss for the $q_1$ is $7.84GeV$, which corresponds to a transformer ratio: $T.R.=0.8$ from the driver to positron bunch. While the loading electron bunch experienced average energy gain $15GeV$ since it sits on a much higher accelerating gradient. Fig.~\ref{Fig4}~(d) shows the $x$ and $y$ component of the normalized transverse emittance, i.e., $\epsilon_{nx}=\sqrt{<x^2><p_x^2>-<xp_x>^2}$. The total emittance of the positron bunch increased $2.1\%$during the whole acceleration process. Another simulation is done by using a $45pC$ positron charge with the same size and at the same location. Without optimizing its current profile and beam loading, we observed a transformer ratio: $T.R.=0.5$ and energy spread $\Delta E/E=10\%$. The unoptimized beam loading from the positron charge also affects the uniformity of the linear focusing structure and that results in emittance growth. However, the beam-loading of positron can be optimized by using a similar technique mentioned in~\cite{BeamLoading_2008} and that will be subject to future studies.

{\it Discussion and Outlook.} 

We have demonstrated that the ELBA scheme is capable of producing robust positron accelerating structure in a nonlinear bubble regime without relying on a special driver~\cite{Hollow_Positron_2015} or a special plasma profile~\cite{Column_Positron_2019}. And the two-electron-beam scenario used in the ELBA can already be produced experimentally, for example, at~\cite{BNL_twoBeam_2008}. Moreover, the ELBA scheme is not restricted to the electron drivers, additional simulation shows that ELBA can be achieved by combining a laser pulse and electron beam within the capacity of the current facility~\cite{BNL_overview} (see Sec. III of Supplemental Material). The ELBA scheme also provides a way to simultaneously accelerate electron and positron bunch in the same regime. We have also shown that the positron accelerating structure only depends on the loading electron bunch and is independent of the driver bunch profile, therefore, high transformer ratio acceleration in the ELBA scheme can be achieved by increasing the driver bunch length~\cite{Lu_TR_2009}. 
Outside the linear focusing region, an area with transverse size $\approx0.8k_p^{-1}$ can still provide focusing to the positron bunch, however, the positron bunch needs to be quasi-matched in that area as described in~\cite{Column_Positron_2020}.  So far, the loading effect of the positron charge is not considered in the simulations shown in this letter, i.e., the influence of the beam loading from positron on the accelerating structure and the flattening of the accelerating field to preserve the bunch quality would be subject to future research.

In Fig.~\ref{Fig2} we showed that the focusing field in a quasi-linear regime is not ideal for emittance preservation due to the betatron decoherency in the longitudinal direction. In fact, this drawback also makes the QL regime less appealing in another way: the betatron decoherency severely damages the guiding of the driver bunch unless the driver bunch is specially crafted slice per slice. Experiments and simulations in the QL and linear regime~\cite{QuasiLinear_2017} show that the energy usage of driver bunch is poor: $\Delta E_{d}[\%]<10\%$ even when the emittance of driver bunch is matched in the plasma. Therefore, although the linear or QL regime may have an advantage in T.R., the energy efficiency on a single positron bunch: $T.R.\times \Delta E_{d}[\%]$, is much less than what in ELBA. For example: in Fig.~\ref{Fig4}, $T.R.=0.8$ and $\Delta E_{d}[\%]=78\%$ since the majority of driver is guided well in a blowout bubble. That's another reason why people would favor a nonlinear regime over a QL regime.

In conclusion, we proposed and demonstrated in simulations, that the ELBA regime would be a promising regime for robust positron accelerations. High-quality, high-efficiency acceleration of positron bunch can be achieved, which paves the pathway to future plasma-based lepton colliders.

\section{ACKNOWLEDGMENT}
This work is supported by the DOE Grant No. DE- SC-0019431. The authors thank the Texas Advanced Computing Center (TACC) at The University of Texas at Austin for providing HPC resources.


\begin{thebibliography}{10}



\bibitem{Colliders1}
	W. Leemans and E. Esarey, 
	"Laser-driven plasma-wave electron accelerators,"
	{\em Phys. Today}, vol.~62, 44 (2009).

\bibitem{Colliders2}
    C. B. Schroeder, E.~Esarey, C. G. R. Geddes, C. Benedetti, and W. P. Leemans,
    "Physics considerations for laser-plasma linear colliders,"
    {\em Phys. Rev. ST Accel. Beams}, vol.~13, 101301 (2010).

\bibitem{RoadMap}
	"Advanced accelerator development strategy report: DOE advanced accelerator concepts research roadmap workshop, Technical Report," 
	U.S. DOE Office of Science, Washington, DC, (2016).

\bibitem{CLIC}
	CLIC collaboration,
	"A multi-TEV linear collider based on CLIC technology: CLIC Conceptual Design Report,"
	CERN, Geneva, Switzerland, Rep. CERN-2012-007 (2012).

\bibitem{InternationalLC}
	T. Behnke, J. E. Brau, B. Foster, J. Fuster, M. Harrison, J. M. Paterson, M. Peskin, M. Stanitzki, N. Walker, and H. Yamamoto,
	"The international linear collider technical design report–Volume 1: Executive summary,"
	arXiv:1306.6327.

\bibitem{LWFA_1}
    V.~Malka, J.~Faure, Y.~A.~Gauduel, E.~Lefebvre, A.~Rousse, and K.~T.~Phuoc,
    "Principles and applications of compact laser–plasma accelerators,"
    {\em Nat. Phys.}, vol.~4, 447 (2008).
\bibitem{LWFA_2}
    E.~Esarey, C.~B.~Schroeder, and W.~P.~Leemans,
    "Physics of laser-driven plasma-based electron accelerators,"
    {\em Rev. Mod. Phys.}, vol.~81, 1229 (2009).

\bibitem{LWFA_3}
    S.~M.~Hooker,
    "Developments in laser-driven plasma accelerators,"
    {\em Nat. Photonics}, vol.~7, 775 (2013).

\bibitem{PWFA_1}
	T. Katsouleas,
	"Physical mechanisms in the plasma wake-field accelerator,"
	{\em Phys. Rev. A}, vol.~33, 2056 (1986).


\bibitem{GeV_0}
    K.~Nakamura, B.~Nagler, C.~Tóth, C.~G.~R.~Geddes, C.~B.~Schroeder, E.~Esarey, S.~M.~Hooker,
    "GeV electron beams from a centimeter-scale channel guided laser wakefield accelerator,"
    {\em Phys. Plasmas}, vol.~14, 056708 (2007).
\bibitem{GeV_1}

    X.~Wang, R.~Zgadzaj, N.~azel, Z.~Li, S.~A.~Yi, X.~Zhang, W.~Henderson, Y.-Y.~Chang, R.~Korzekwa, H.-E.~Tsai, C.-H. ~Pai, H.~Quevedo, G.~Dyer, E.~Gaul, M.~Martinez, A.~C.~Bernstein, T.~Borger, M.~Spinks, M.~Donovan, V.~Khudik, G.~Shvets, T.~Ditmire and M.~C.~Downer,  
    "Quasi-monoenergetic laser-plasma acceleration of electrons to 2 GeV,"
    {\em Nat. Comms.}, vol.~4, 1988 (2013).

\bibitem{GeV_2}
    W.~P.~Leemans, A.~J.~Gonsalves, H.-S. Mao, K.~Nakamura, C.~Benedetti, C.~B.~Schroeder, Cs.~Tóth, J. Daniels, D.~E.~Mittelberger, S.~S.~Bulanov, J.-L.~Vay, C.~G.~R.~Geddes, and E.~Esarey,
    "Multi-GeV electron beams from capillary-discharge-guided subpetawatt laser pulses in the self-trapping regime,"
    {\em Phys. Rev. Lett.}, vol.~113 245002 (2014).

\bibitem{GeV_3}
    H.~T.~Kim, V. B. Pathak, K. H. Pae, A. Lifschitz, F. Sylla, J. H. Shin,  C. Hojbota, S. Ku. Lee, J. H. Sung, H. W. Lee, E. Guillaume, C. Thaury, K. Nakajima, J. Vieira, L. O. Silva, V. Malka and C. H. Nam,
    "Stable multi-GeV electron accelerator driven by waveform-controlled PW laser pulses,"
    {\em Sci. Rep.}, vol.~7, 10203 (2017).

\bibitem{GeV_2b}
    A. J. Gonsalves, K. Nakamura, J. Daniels, C. Benedetti, C. Pieronek, T. C. H. de Raadt, S. Steinke, J. H. Bin, S. S. Bulanov, J. van Tilborg, C. G. R. Geddes, C. B. Schroeder, Cs. Tóth, E. Esarey, K. Swanson, L. Fan-Chiang, G. Bagdasarov, N. Bobrova, V. Gasilov, G. Korn, P. Sasorov, and W. P. Leemans,
    "Petawatt Laser Guiding and Electron Beam Acceleration to 8 GeV in a Laser-Heated Capillary Discharge Waveguide,"
    {\em Phys. Rev. Lett.}, vol.~122, 084801 (2019).

\bibitem{PWFA_GeV_1}
	M. J. Hogan, C. D. Barnes, C. E. Clayton, F. J. Decker, S. Deng, P. Emma, C. Huang, R. H. Iverson, D. K. Johnson, C. Joshi, T. Katsouleas, P. Krejcik, W. Lu, K. A. Marsh, W. B. Mori, P. Muggli, C. L. O’Connell, E. Oz, R. H. Siemann, and D. Walz,
	"Multi-GeV Energy Gain in a Plasma-Wakefield Accelerator,"
	{\em Phys. Rev. Lett.}, vol.~95, 054802 (2005).

\bibitem{PWFA_GeV_2}
	M. Litos, E. Adli, W. An, C. I. Clarke, C. E. Clayton, S. Corde, J. P. Delahaye, R. J. England, A. S. Fisher, J. Frederico, S. Gessner, S. Z. Green, M. J. Hogan, C. Joshi, W. Lu, K. A. Marsh, W. B. Mori, P. Muggli, N. Vafaei-Najafabadi, D. Walz, G. White, Z. Wu, V. Yakimenko, and G. Yocky,
	"High-efficiency acceleration of an electron beam in a plasma wakefield accelerator,"
	{\em Nature}, vol.~515, 92 (2014).

\bibitem{PWFA_GeV_3}
	I. Blumenfeld, C. E. Clayton, F. J. Decker, M. J. Hogan, C. Huang, R. Ischebeck, R. Iverson, C. Joshi, T. Katsouleas, N. Kirby, W. Lu, K. A. Marsh, W. B. Mori, P. Muggli, E. Oz, R. H. Siemann, D. Walz, and M. Zhou,
	"Energy doubling of 42 GeV electrons in a metre-scale plasma wakefield accelerator,"
	{\em Nature}, vol.~445, 741, (2007).

\bibitem{PWFA_Field_1991}
	J. B. Rosenzweig, B. Breizman, T. Katsouleas, and J. J. Su,
	"Acceleration and focusing of electrons in two-dimensional nonlinear plasma wake fields,"
	{\em Phys. Rev. A}, vol.~44, R6189 (1991).

\bibitem{LWFA_Field_2002}
	A. Pukhov, and J. Meyer-ter-Vehn, 
	"Laser wake field acceleration: the highly non-linear broken-wave regime,"
	{\em Appl. Phys. B}, vol.~74, 355 (2002). 

\bibitem{Mapping_Field_2016}
	C. E. Clayton, E. Adli, J. Allen, W. An, C. I. Clarke, S. Corde, J. Frederico, S. Gessner, S. Z. Green, M. J. Hogan, C. Joshi, M. Litos, W. Lu, K. A. Marsh, W. B. Mori, N. Vafaei-Najafabadi, X. Xu, and V. Yakimenko,
	"Self-mapping the longitudinal field structure of a nonlinear plasma accelerator cavity,"
	{\em Nat. Comm.}, vol.~7, 12483 (2016).

\bibitem{Lotov_Positron}
	K. V. Lotov,
	"Acceleration of positrons by electron beam-driven wakefields in a plasma,"
	{\em Phys. Plasmas}, vol.~14, 023101 (2007).

\bibitem{Nature_Positron_2015}
    S. Corde, E. Adli, J. Allen, W. An, C. Clarke, C. Clayton, J. Delahaye, J. Frederico, S. Gessner, S. Green et al.,
    "Multi- gigaelectronvolt acceleration of positrons in a self-loaded plasma wakefield,"
    {\em Nature}, vol.~524, 442 (2015).

\bibitem{Hollow_Positron_2015}
    N. Jain, T. M. Antonsen, Jr., and J. P. Palastro,
    "Positron Acceleration by Plasma Wakefields Driven by a Hollow Electron Beam,"
    {\em Phys. Rev. Lett.}, vol.~115, 195001 (2015).

\bibitem{Column_Positron_2019}
    S. Diederichs, T. J. Mehrling, C. Benedetti, C. B. Schroeder, A. Knetsch, E. Esarey, and J. Osterhoff,
    "Positron transport and acceleration in beam-driven plasma wakefield accelerators using plasma columns,"
    {\em Phys. Rev. Accel. Beams}, vol.~22, 081301 (2019).

\bibitem{Column_Positron_2020}
    S. Diederichs, C. Benedetti, E. Esarey, J. Osterhoff, and C. B. Schroeder,
    "High-quality positron acceleration in beam-driven plasma accelerators,"
    {\em Phys. Rev. Accel. Beams}, vol.~23, 121301 (2020).


\bibitem{stupakov_2016}
    G.~Stupakov, B.~Breizman, V.~Khudik, and G.~Shvets,
    "Wake excited in plasma by an ultrarelativistic pointlike bunch,"
    {\em Phys. Rev. Accel. Beams}, vol.~19, 101302 (2016).

\bibitem{My_Driver_2017}
    T.~Wang, V.~Khudik, B.~Breizman, and G.~Shvets,
    "Nonlinear plasma waves driven by short ultrarelativistic electron bunches,"
    {\em J. Plasma Phys.} vol.~24, 103117 (2017).

\bibitem{WAND2020}
    T.~Wang, V.~Khudik, and G.~Shvets,
    "WAND-PIC: A three-dimensional quasi-static particle-in-cell code with parallel multigrid solver and without predictor-corrector,"
    arXiv:2012.00881 (2020).

\bibitem{VLPL1999}
    A. Pukhov,
    "Three-dimensional Electromagnetic Relativistic Particle-in-cell Code VLPL (Virtual Laser Plasma Lab),"
    {\em J. Plasma Phys.} vol.~61, 425 (1999).

\bibitem{VLPL2020}
    A. Pukhov,
    "X-dispersionless Maxwell solver for plasma-based particle acceleration," 
    {\em J. Comp. Phys.} vol.~418 (2020).

\bibitem{FACETII_2018}
	C. Joshi, E. Adli, W. An, C. E. Clayton, S. Corde, S. Gessner, M. J. Hogan, M. Litos, W. Lu, K. A. Marsh, W. B. Mori, N. Vafaei-Najafabadi, B. O'shea, Xinlu Xu, G. White, and V. Yakimenko,
	"Plasma wakefield acceleration experiments at FACET II,"
	 {\em Plasma Phys. Controlled Fusion}, vol.~60, 034001 (2018).

\bibitem{Emittance_Mehrling_2012}
	T. Mehrling, J. Grebenyuk, F. S. Tsung, K. Floettmann, and J. Osterhoff,
	"Transverse emittance growth in staged laser-wakefield acceleration,"
	{\em Phys. Rev. ST Accel. Beams} vol.~15, 111303 (2012).

\bibitem{Emittance_Xu_2012}
	X. L. Xu, J. F. Hua, F. Li, C. J. Zhang, L. X. Yan, Y. C. Du, W. H. Huang, H. B. Chen, C. X. Tang, W. Lu, P. Yu, W. An, C. Joshi, and W. B. Mori,
	"Phase-Space Dynamics of Ionization Injection in Plasma-Based Accelerators,"
	{\em Phys. Rev. Lett.} vol.~112, 035003  (2014).

\bibitem{QuasiLinear_2017}
	A. Doche, C. Beekman, S. Corde, J. M. Allen, C. I. Clarke, J. Frederico, S. J. Gessner, S. Z. Green, M. J. Hogan, B. O’Shea, V. Yakimenko, W. An, C. E. Clayton, C. Joshi, K. A. Marsh, W. B. Mori, N. Vafaei-Najafabadi, M. D. Litos, E. Adli, C. A. Lindstrøm, and W. Lu,
	"Acceleration of a trailing positron bunch in a plasma wakefield accelerator,"
	{\em Sci. Rep.} vol.~7, 14180 (2017).

\bibitem{QL_An_2019}
	W. An,
	"Positron Acceleration in the Electron Driven Plasma Wake Field,"
	{\em In ALEGRO Positron Acceleration in Plasma Mini-Workshop} (2018).

\bibitem{WB_JETP_1956}
	A. I. Akhiezer and R. V. Polovin, 
	“Theory of Wave Motion of an Electron Plasma,” 
	{\em Soviet Phys. JETP} vol.~3, 696 (1956).

\bibitem{WB_Dawson_1959}
	John M. Dawson,
	"Nonlinear Electron Oscillations in a Cold Plasma,"
	{\em Phys. Rev.} vol.~113 383 (1959).

\bibitem{Esarey_Temp_2004}
	B. A. Shadwick, G. M. Tarkenton, and E. H. Esarey,
	"Hamiltonian Description of Low-Temperature Relativistic Plasmas,"
	{\em Phys. Rev. Lett.} vol.~93 175002 (2004).

\bibitem{Esarey_Temp_2005}
	B. A. Shadwick, G. M. Tarkenton, E. Esarey, and C. B. Schroeder,
	"Fluid and Vlasov models of low-temperature, collisionless, relativistic plasma interactions,"
	{\em Phys. Plasmas} vol.~12, 056710 (2005)


\bibitem{WB_Warm_Coffey}
	T. P. Coffey,
	"Breaking of Large Amplitude Plasma Oscillations,"
	{\em Phys. Fluids} vol.~14, 1402 (1971)

\bibitem{WB_Warm_Schroeder}
	C. B. Schroeder, E. Esarey, and B. A. Shadwick,
	"Warm wave breaking of nonlinear plasma waves with arbitrary phase velocities,"
	{\em Phys. Rev. E} vol.~72, 055401 (2005).

\bibitem{WB_Warm_Schroeder2}
	C. B. Schroeder and E. Esarey,
	"Relativistic warm plasma theory of nonlinear laser-driven electron plasma waves,"
	{\em Phys. Rev. E}, vol.~81, 056403 (2010).

\bibitem{Channel_ionization}
	C. G. Durfee, III, J. Lynch, and H. M. Milchberg,
	"Development of a plasma waveguide for high-intensity laser pulses,"
	{\em Phys. Rev. E} vol.~51 (1995).

\bibitem{Channel_ionization2}
	H. M. Milchberg, T. R. Clark, C. G. Durfee III, and T. M. Antonsen,
	"Development and applications of a plasma waveguide for intense laser pulses,"
	{\em Phys. Plasmas}, vol.~3 (1996).

\bibitem{Lu_bubble_2006}
	W. Lu, C. Huang, M. Zhou, W. B. Mori, and T. Katsouleas,
	"Nonlinear Theory for Relativistic Plasma Wakefields in the Blowout Regime,"
	{\em Phys. Rev. Lett.}, vol.~96, 165002 (2006).

\bibitem{Yi_bubble_2013}
	S. A. Yi, V. Khudik, C. Siemon, and G. Shvets,
	"Analytic model of electromagnetic fields around a plasma bubble in the blowout regime,"
	{\em Phys. Plasmas} vol.~20, 013108 (2013).

\bibitem{BeamLoading_2008}
	M. Tzoufras, W. Lu, F. S. Tsung, C. Huang, W. B. Mori, T. Katsouleas, J. Vieira, R. A. Fonseca, and L. O. Silva,
	"Beam Loading in the Nonlinear Regime of Plasma-Based Acceleration,"
	{\em Phys. Rev. Lett.} vol.~101, 145002 (2008).

\bibitem{BNL_twoBeam_2008}
	E. Kallos, T. Katsouleas, W. D. Kimura, K. Kusche, P. Muggli, I. Pavlishin, I. Pogorelsky, D. Stolyarov, and V. Yakimenko,
	"High-Gradient Plasma-Wakefield Acceleration with Two Subpicosecond Electron Bunches,"
	{\em Phys. Rev. Lett.} vol.~100, 074802 (2008).

\bibitem{BNL_overview}
	I. V. Pogorelsky, and I. Ben-Zvi,
	"Brookhaven National Laboratory's Accelerator Test Facility: research highlights and plans,"
	{\em Plasma Phys. Controlled Fusion} vol.~56, 084017 (2014).

\bibitem{Lu_TR_2009}
	W. Lu, W. An, C. Huang, C. Joshi, W. B. Mori, M. Hogan, T. Raubenheimer, A. Seryi, P. Muggli, and T. Katsouleas,
	"High transformer ratio PWFA for application on XFELs,"
	{\em Proc. PAC09}, 3028 (2009).  




\end{thebibliography}
\end{document}